\documentclass[journal]{IEEEtran}  % journal,   %onecolumn,12pt,draftcls
\usepackage[margin=1in,bottom=1.1in]{geometry} 
\normalsize

\usepackage{cite}
\usepackage[pdftex]{graphicx}
\usepackage[cmex10]{amsmath}
\usepackage{amssymb}
\usepackage{color}
\usepackage{multirow}
\usepackage{algpseudocode}
\usepackage[]{algorithm}
\usepackage{setspace}
\usepackage{subcaption}
\usepackage[font=footnotesize]{caption}
\usepackage[font=footnotesize]{subcaption}

\newtheorem{proposition}{Proposition}

\usepackage{soul}

\newcommand{\rom}[1]{\uppercase\expandafter{\romannumeral #1\relax}}

\title{Discrete Dynamic Causal Modeling and Its Relationship with Directed Information}

\author{Zhe~Wang,
		Yu~Zheng,
        David~C.~Zhu, %,~\IEEEmembership{}
        Jian~Ren
        and~Tongtong~Li%,~\IEEEmembership{}% <-this % stops a space

	\thanks{Zhe Wang, Yu Zheng, Jian Ren and Tongtong Li are with the Department of Electrical and Computer Engineering; David C. Zhu is with the Department of Radiology, Michigan State University, East Lansing, MI, 48824, USA (e-mail: wangzh34@msu.edu, zhengy30@msu.edu, zhuda@msu.edu, renjian@egr.msu.edu, tongli@egr.msu.edu)}% <-this % stops a space
}% <-this % stops a spaces

\begin{document}
\maketitle

%=================================================================================================
% Abstract and Keywords
%=================================================================================================
\begin{abstract}

%\vskip -0.15in

This paper explores the discrete Dynamic Causal Modeling (DDCM) and its relationship with Directed Information (DI). We prove the conditional equivalence between DDCM and DI in characterizing the causal relationship between two brain regions. The theoretical results are demonstrated using fMRI data obtained under both resting state and stimulus based state. Our numerical analysis is consistent with that reported in previous study.

\end{abstract}


\IEEEpeerreviewmaketitle

% ===============================================================

\section{Introduction}
Causality analysis  aims to find the relationship between causes and effects. It provides insightful information  on how brain regions interact with each other during a cognitive task~\cite{Causal_fmri}. %In general, causality analysis tries to determine whether the values of one time series are useful in predicting the future values of another time series. Since 1990s, a number of frameworks have been applied to fMRI based causality analysis. 
In fMRI based causality analysis, \textit{Granger Causality} (GC), \textit{Directed Information} (DI), and \textit{Dynamic Causal Modeling} (DCM) are three representative approaches. In this paper, we will revisit the relationship among existing  causality analysis frameworks,  especially the relationship between DCM and DI.

\emph{Granger causality}  is the  first practical causality analysis framework which was proposed by Granger in 1969~\cite{Granger}. The basic idea is, if two signals  $X_1$ and $X_2$ form a causal relationship, then, instead of using the past values of $X_2$ alone, the information contained in the past values of $X_1$ will help predict $X_2$. In practice, the calculation of Granger Causality is based on the linear prediction models.

In literature, there have been growing interests in applying GC to identify causal interactions in the brain~\cite{Friston,hu2014copula,Hu,David_Granger}. As a widely accepted technique, the validity and computational simplicity of Granger Causality have been appreciated. However, it has also been noticed that GC relies heavily on the linear prediction method. When there exist instantaneous and/or strong nonlinear interactions between two regions, GC analysis may lead to invalid results~\cite{David_Granger,di}.

\emph{Directed Information} is an information theoretic metric, which was first introduced by Massey when studying communication channels with feedback~\cite{Massey}. It measures the directed information flow from one time series $X$ to another time series $Y$, denoted as $I(X\rightarrow Y)$. If $I(X\rightarrow Y) > I(Y\rightarrow X)$, then we say $X$ has more influence on $Y$, or $X$ is the causal side in the connectivity. 

DI can be used to characterize more general relationships, as it does not have	any modeling constraints on the sequences to be evaluated.  In~\cite{di}, it was pointed out that GC analysis is effective in detecting linear or nearly linear causal relationship, but may have difficulty in capturing nonlinear causal relationships. On the other hand, DI-based causality analysis is more effective in capturing both linear and nonlinear causal relationships. Moreover, in~\cite{Pierre}, it was shown that the Granger Causality graphs of stochastic processes can be generated from the DI framework, and the authors indicated that DI provides an effective framework for connectivity inference. % A comprehensive investigation of DI can be found in \cite{Weissman}. 

\emph{Dynamic Causal Modeling}, which characterizes the interactions among a group of brain regions~\cite{Friston_dcm}, was proposed by Friston in 2003. It assumes that the invisible neurostate, the (external) input $U$, and the observed BOLD signal $Y$ form a dynamic system that could be described by a group of differential equations.

%\textbf{Dynamic Causal Modeling} In 2003, Friston proposed the framework of \textit{Dynamic Causal Modeling} to characterize the interactions among a group of brain regions~\cite{Friston_dcm}. DCM assumes that the invisible neurostate $X$, the observed BOLD signal $Y$, the (external) input $U$, the parameter $\theta$ that characterizes the connectivities between different brain regions, and the independent noise $\Omega$ form a dynamic system that could be described by the following equations: 
%\begin{eqnarray}
%\dot{X}  =  f(X,U,\theta) \mbox{~and}~ Y  =  \Lambda(X) + \Omega,
%\end{eqnarray}
%where $\Lambda$ represents the mapping from the neurostate $X$ to the observed BOLD signal $Y$. Relying on the EM algorithm, DCM has been implemented on both fMRI and EEG data~\cite{Stephan_dcm}. In practical applications, due to the computational complexity,  DCM is usually used as a confirmatory approach. That is, the users need to put forward different connectivity models and then compare them based on their likelihood evaluated under DCM~\cite{Ryali}. 

Compared with GC, DCM provides a more comprehensive characterization of the dynamic interactions between multiple regions. In~\cite{friston_gc_dcm}, Friston et al. pointed out that GC and DCM were complementary to each other: GC models the causal dependency among observed responses, while DCM models the causal interactions among the hidden neurostates. On the other hand, in~\cite{friston_dcm_gc}, Friston provided an example to show  that DCM and GC may generate different results given the same dataset. The underlining argument is that: DCM takes into account both the external input and the biological variations of the hemodynamic response, which are not involved in GC.

To this end, it can be seen that the relationships between GC and DCM, and between GC and DI have been investigated in literature. While GC is efficient in detecting linear causal relationships, both DI and DCM can be used to characterize more general causal relationships. A missing link here is: what is the relationship between DCM and DI?

In the following sections, we aim to fill this missing link, and explore the connection between DCM and DI. Based on the discrete DCM (DDCM), we will show that under certain conditions,  DDCM and DI are equivalent in characterizing the causal relationship between two brain regions. This equivalence is validated using fMRI data obtained under both resting state and stimulus based state.

%The rest of this paper is organized as follows. In Section \rom{2}, we revisit DDCM and demonstrate its relationship with the continuous time DCM. In Section \rom{3}, we prove the conditional equivalence between DDCM and DI in characterizing the causal relationship between two brain regions. We present the numerical results in Section \rom{4}, and conclude in Section \rom{5}.

\section{Discrete Dynamic Causal Modeling}

In continuous time DCM, the invisible neurostates of $d$ brain regions are denoted by a vector $\mathbf{X} = [X_1, X_2, ..., X_d]^t$, where each $X_i$, $i=1,2,...,d$, represents the neurostate of the $i$th region. The basic idea of DCM is that, the neurostate $\mathbf{X}$, the external input $U$, the connectivity matrices $\tilde{A}$ and $\tilde{B}$ that describe the connections among brain regions, the observed BOLD signal $\mathbf{Y}$ and the independent noise can be formulated as a complex dynamic system, characterized as:
\begin{eqnarray}
\dot{\mathbf{X}}(t) & = & \tilde{A}\mathbf{X}(t) + \tilde{B}U(t) + \mathbf{\Omega}_1(t),  \label{eq:dcm_def_1} \\
\mathbf{Y}(t) & = & \tilde{\Lambda}(\mathbf{X}(t))  + \mathbf{\Omega}_2(t), \label{eq:dcm_def_2}
\end{eqnarray}
where $\mathbf{\Omega}_1(t)$ and $\mathbf{\Omega}_2(t)$ are the state noise and observation noise, and $\tilde{\Lambda}$ represents the mapping from the neurostate $\mathbf{X}(t)$ to the observed BOLD signal $\mathbf{Y}(t)$.

%The functional connectivity in DCM is mainly characterized by matrix $\tilde{A}$~\cite{Friston}. For instance, in a model with two brain regions, that is, $d = 2$ and $\mathbf{X} = [X_1, X_2]^t$, the connectivity matrix $\tilde{A}$ will be:
%\begin{equation}
%\tilde{A} = \left[\begin{array}{cc}
%\tilde{A}_{11} & \tilde{A}_{12} \\
%\tilde{A}_{21} & \tilde{A}_{22}
%\end{array}
%\right]. \label{eq:dcm_A}
%\end{equation}
%Here, for $i, j = 1,2$, $\tilde{A}_{ii}$ measures the influence of the past values of $X_i$ on its future values, and $\tilde{A}_{ji}$ measures the influence of past values of $X_i$ on the future values of $X_j$. The absolute values of $\tilde{A}_{12}$ and $ \tilde{A}_{21}$ describe the causal relationship between the two regions: when $|\tilde{A}_{12}| > |\tilde{A}_{21}|$, it means that $X_2$ has imposed more influence over $X_1$; and when $|\tilde{A}_{21}| > |\tilde{A}_{12}|$, it means that $X_1$ has imposed more influence over $X_2$~\cite{Smith_switch_dcm,Ryali}.

It can be seen from equations (\ref{eq:dcm_def_1}) and (\ref{eq:dcm_def_2}) that the continuous time DCM characterizes the dynamic neural system using two continuous-time  equations. However, parameter  estimation in continuous time equations faces considerable challenges in practical applications. To overcome this difficulty, there have been efforts to simplify DCM to a more tractable form, such as the switching linear dynamic model (SLDS)~\cite{Smith_switch_dcm} and  multivariate dynamical model (MDS)~\cite{Ryali}. In both approaches, the continuous time equations are discretized, and the mapping between the neurostate and the BOLD signal is approximated as an LTI system, characterized using a convolution. That is, the discrete DCM (DDCM) model can be obtained as:
\begin{eqnarray}
\mathbf{X}(k+1)		& = & A\mathbf{X}(k) + BU(k) + \mathbf{\Omega}_1(k),	\label{eq:sdcm_state} \\
\mathbf{Y}(k)		& = & \sum_{m=0}^{M}\Lambda(m)\mathbf{X}(k-m) + \mathbf{\Omega}_2(k), \label{eq:sdcm_obs}
\end{eqnarray}
where $A$ is the  connectivity matrix, $\{\Lambda(m), m=0,1,\cdots M\}$  denotes the convolution coefficients corresponding to the hemodynamic response, and $\mathbf{\Omega}_1(k)$ and $\mathbf{\Omega}_2(k)$ denote the noise terms independent of the brain state and the input.

Consider the case of two regions, region $1$ and region $2$, where equation (\ref{eq:sdcm_state}) can be rewritten as:
\begin{eqnarray}
\left[\begin{array}{c}
X_1(k+1) \\
X_2(k+1) \end{array}\right] & = &
\left[\begin{array}{cc}
A_{11} & A_{12} \\
A_{21} & A_{22}
\end{array}
\right] \left[\begin{array}{c}
X_1(k) \\
X_2(k)
\end{array}
\right] \nonumber\\ & + & \left[\begin{array}{c}
B_1 \\
B_2
\end{array}
\right] U(k) + \left[\begin{array}{c}
\Omega_{11}(k) \\
\Omega_{12}(k)
\end{array}
\right].	\IEEEeqnarraynumspace				\label{eq:sdcm_two_detail}
\end{eqnarray}

Similar to the continuous time DCM, coefficients $A_{12}$ and $A_{21}$ actually measure the causal relationship between region $1$ and region $2$. More specifically, if $|A_{21}| > |A_{12}|$,  then $X_1$ is more likely to be the casual side, and vice versa. The same analysis holds when multiple brain regions are under investigations~\cite{Smith_switch_dcm,Ryali}.

%In the following sections, we will investigate the relationship of DDCM with the DI-based causality analysis framework.

\section{The Relationship between DDCM and Directed Information}

In this section, we show that, under certain assumptions, DI and DDCM  are equivalent in characterizing the causal relationship between two brain regions.

Directed Information is a causality analysis framework based on information theory. 
The directed information from one time series $\textbf{X}_1^n$ to another $\textbf{X}_2^n$ is calculated as~\cite{Massey}:
\begin{IEEEeqnarray}{Rl}
&I(\textbf{X}_1^n\rightarrow \textbf{X}_2^n) \IEEEnonumber\\
&= \sum_{k=1}^{n}[h(X_2(k)|\textbf{X}_2^{k-1}) - h(X_2(k)|\textbf{X}_2^{k-1},\textbf{X}_1^k)]  \label{eq: di-def},
\end{IEEEeqnarray}
where $\textbf{X}_i^k = [X_i(1),X_i(2),...,X_i(k)]$, $i = 1,2$, and $h$ denotes the differential entropy operator.  If $I(\textbf{X}_1^n\rightarrow \textbf{X}_2^n)$ is  greater than $I(\textbf{X}_2^n\rightarrow \textbf{X}_1^n)$, we say $\textbf{X}_1^n$ has more causal influence over $\textbf{X}_2^n$; otherwise $\textbf{X}_2^n$ has more causal influence over $\textbf{X}_1^n$.

%\subsection{DDCM and Directed Information}

When deriving the relationship between DDCM and DI, we impose the following assumptions to make the problem more tractable: (i) The dynamic neural system under investigation is a causal system, which means for each brain region, the current value of the neurostate depends only on previous values of neurostates of the region and its related regions. (ii) For each region, both the neurostate and the background noise are normally distributed, and the variances are the same in related brain regions. More specifically, for each $k = 1, 2, \cdots,n$, the variances corresponding to the neurostate and the background noise are $\sigma_x^2$ and $\sigma_0^2$, respectively. (iii) The external input $U$ is a constant.  This assumption is reasonable when the changing rate of the external input is much slower than that of neurostates.

In the following analysis, let the uppercase letters ($X$,$Y$,...) denote random variables, and the lowercase letters ($x$,$y$,...) the possible values they can acquire. In particular, $x_1(k)$ and $x_2(k)$ denote the possible values $X_1(k)$ and $X_2(k)$ can acquire, and $\omega_{11}(k)$ and $\omega_{12}(k)$ denote the possible values $\Omega_{11}(k)$ and $\Omega_{12}(k)$ can acquire.  Given a time series $\textbf{X}^n = [X(1),X(2),...,X(n)], n \in N$, for any $x(k), k=1,2,\dots n$, $ P(x(k))$ denotes the probability for $X(k)$ to take the value $x(k)$, and $ P(x(k)|\textbf{x}^{k-1})$ the conditional probability that the current sample $X(k)$ is $x(k)$, given that the previously observed sequence is $\textbf{x}^{k-1}=[x(1),x(2),...,x(k-1)] $.

Following  equation (\ref{eq:sdcm_two_detail}), the conditional probability $P(x_2(k) ~|~ \textbf{x}_2^{k-1},\textbf{x}_1^k)$ can be written as:
\begin{IEEEeqnarray}{Rl}
	&P(x_2(k) ~|~ \textbf{x}_2^{k-1},\textbf{x}_1^k) \IEEEnonumber\\
	=&P(A_{21}x_1(k-1)+A_{22}x_2(k-1)  \IEEEnonumber\\
	&+\: B_2U+\omega_{12}(k-1) ~|~ \textbf{x}_2^{k-1},\textbf{x}_1^{k}) \IEEEnonumber\\
    =& P(A_{21}x_1(k-1)+A_{22}x_2(k-1) \IEEEnonumber\\
	&+\: B_2U+\omega_{12}(k-1) ~|~ \textbf{x}_2^{k-1},\textbf{x}_1^{k-1}) \IEEEnonumber\\
    =& P(\omega_{12}(k-1)).
	\label{eq:dcm_simplify_prob}
\end{IEEEeqnarray}
This implies that, for each $k = 1, 2, \cdots,n$, the conditional probability density function of the neurostate $X_2(k)$ given $\textbf{X}_2^{k-1}$ and $\textbf{X}_1^k$ is Gaussian with variance $\sigma_0^2$.

It is well known that given a Gaussian random variable $\Xi$ with variance $\sigma_{\xi}^2$, the corresponding differential entropy $h(\Xi)$ can be calculated as:
\begin{eqnarray}
h(\Xi) = {1\over 2}\log2\pi e\sigma_{\xi}^2.  \label{eq:dif_ep}
\end{eqnarray}
Therefore, based on equations (\ref{eq:dcm_simplify_prob}) and (\ref{eq:dif_ep}), the differential entropy corresponding to the neurostate $X_2(k)$ given $\textbf{X}_2^{k-1}$ and $\textbf{X}_1^k$  can then be calculated as: 
\begin{eqnarray}
h(X_2(k)|\textbf{X}_2^{k-1},\textbf{X}_1^k) = {1\over 2}\log2\pi e\sigma_0^2. \label{eq:de_1}
\end{eqnarray}

Similarly, the conditional probability $P(x_2(k) ~|~ \textbf{x}_2^{k-1})$ can be simplified as:
\begin{IEEEeqnarray}{Rl}
&P(x_2(k) ~|~ \textbf{x}_2^{k-1}) \IEEEnonumber\\
=& P(A_{21}x_1(k-1)+A_{22}x_2(k-1)+B_2U \IEEEnonumber\\
& +\omega_{12}(k-1) ~|~ \textbf{x}_2^{k-1}) \IEEEnonumber\\
=& P(A_{21}x_1(k-1)+\omega_{12}(k-1)).		\label{eq:dcm_simplify_entropy}
\end{IEEEeqnarray}
As a result, the corresponding differential entropy will be:
\begin{eqnarray}
h(X_2(k)|\textbf{X}_2^{k-1}) = {1\over 2}\log2\pi e(A_{21}^2\sigma_x^2+\sigma_0^2), \label{eq:de_2}
\end{eqnarray}
where $\sigma_x^2$ is the variance of the neurostate, which is assumed to have no significant changes among related regions and within the observation frame. 

Based on equations (\ref{eq:dcm_simplify_prob}) to (\ref{eq:de_2}), the directed information can then be obtained as:
\begin{eqnarray}
& & I(\textbf{X}_1^n\rightarrow \textbf{X}_2^n) \nonumber \\
& & =  \sum\limits_{k=1}^{n} [ {1\over 2}\log2\pi e(A_{21}^2\sigma_x^2+\sigma_0^2) - {1\over 2}\log2\pi e\sigma_0^2 ] \nonumber \\
& & = {n\over 2}\log(1+A_{21}^2\frac{\sigma_x^2}{\sigma_0^2}) \nonumber \\
& & = {n\over 2}\log(1+cA_{21}^2), \label{eq:dcm-di}
\end{eqnarray}
where $c={\sigma_x^2}/{\sigma_0^2}$ is the ratio of the power of neural activities and the noise power. Similarly, we can prove that  $I(\textbf{X}_2^n\rightarrow \textbf{X}_1^n) = (n/2)log(1+cA_{12}^2)$.

%It can be seen from equation (\ref{eq:dcm-di}) that  after the parameters in DDCM have been obtained, the directed information from one region to another can be calculated accordingly. That is, equation (\ref{eq:dcm-di}) provides an effective method for the estimation of DI.

Note that when $c > 0$, $log(1+cx^2)$ is a monotonically increasing function. Based on the discussions above, we can obtain the following proposition:
\begin{proposition}\label{pro:1}
If  $|A_{21}|> |A_{12}|$, then $I(\textbf{X}_1^n\rightarrow \textbf{X}_2^n) > I(\textbf{X}_2^n\rightarrow \textbf{X}_1^n)$, that is, region $1$ is more likely to be the causal side; otherwise, we will have $ I(\textbf{X}_2^n\rightarrow \textbf{X}_1^n) > I(\textbf{X}_1^n\rightarrow \textbf{X}_2^n)$, and region $2$ is more likely to be the causal side.
\end{proposition}

Proposition \ref{pro:1} is in accordance with the previous analysis for DDCM: the causality analysis can be carried out based on the absolute values of the coefficients of the connectivity matrix.
This means, for a causal dynamic neural system with a constant external input, when the neurostate and the background noise are normally distributed, DI and DDCM are equivalent in characterizing the causal relationship between two brain regions.

On the other hand, in practice, we can only observe the BOLD signal $\textbf{Y}^n$ rather than the neurostate $\textbf{X}^n$. That is, given two brain regions, region 1 and region 2, the calculation of DI can only be carried out on the observations $\textbf{Y}_1^n$ and $\textbf{Y}_2^n$ rather than the neurostates $\textbf{X}_1^n$ and $\textbf{X}_2^n$. However, it can be shown that as long as the hemodynamic system is invertible,  DI calculated using the estimated neurostates is equal to the DI calculated using the observed signals. That is, DDCM and DI are still equivalent in characterizing the causal relationship between brain regions.
\section{Numerical Analysis}

In this section, we briefly describe how to validate the equivalence of DDCM and DI between two regions using  experimental fMRI data obtained under both resting state and stimulus based state.

\subsection{Data Acquisition}

Fourteen right-handed healthy college students (7 males, $23.4 \pm 4.2$ years of age) from Michigan State University volunteered to participate in this study. The experiment was conducted on a 3T GE Signa HDx MR scanner (GE Healthcare, Waukesha, WI) with an 8-channel head coil. 

For each subject, fMRI datasets were collected on a visual stimulation condition with a scene-object fMRI paradigm and then on a resting-state condition. On the visual stimulation fMRI condition, each volume of images were acquired 192 times (8 \textit{min}) while each subject was presented with 12 blocks of visual stimulation after an initial 10 s ``resting" period. In a predefined randomized order, the scenery pictures were presented in 6 blocks and the object pictures were presented in other 6 blocks. In each block, 10 pictures were presented continuously for 25 s (2.5 s for each picture), followed with a 15 s baseline condition (a white screen with a black fixation cross at the center). The subject needed to press his/her right index finger once when the screen was switched from the baseline to picture condition. Detailed experiment setting and procedures of data processing can be found in~\cite{di}.

We test the robustness of our causality analysis techniques against some expected outcomes: under the stimulation fMRI paradigm, the primary visual cortex (V1) and nearby regions are activated first, followed with activation in the parahippocampal place area (PPA) for higher level scene processing. Some but relatively small activations in the left sensorimotor cortex (SMC) is also expected following V1 activations. Under the resting-state condition, neuronal activity is not expected to occur in a sequential manner among above regions. 

\begin{figure}[htbp]
        %\centering
        \begin{subfigure}[b]{0.5\textwidth}
        		\centering
                \includegraphics[width=0.75\textwidth]{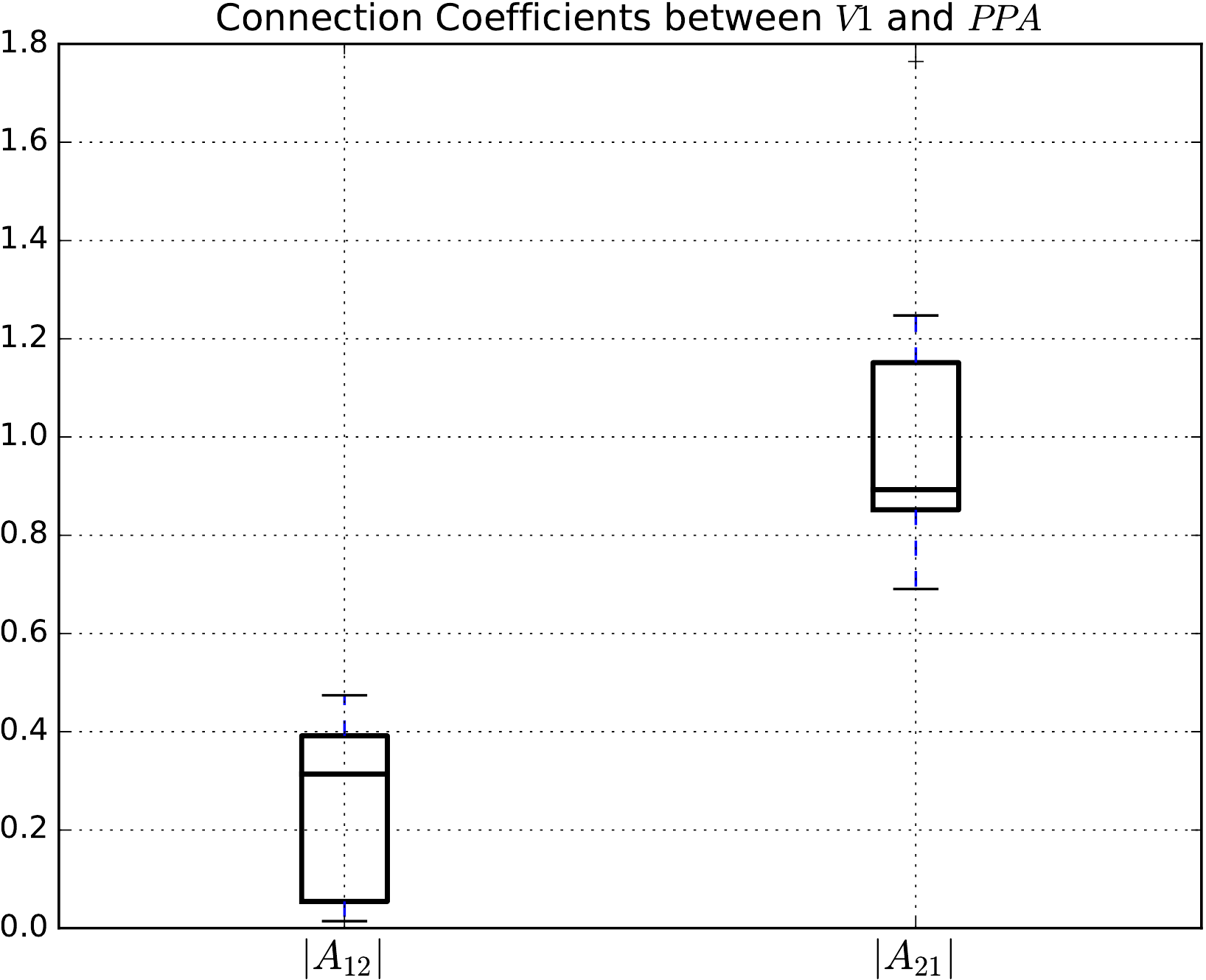}
                \caption{V1 and PPA under stimulus based state.}
                \label{fig:vp}
        \end{subfigure}%
        
        \begin{subfigure}[b]{0.5\textwidth}
        		\centering
                \includegraphics[width=0.75\textwidth]{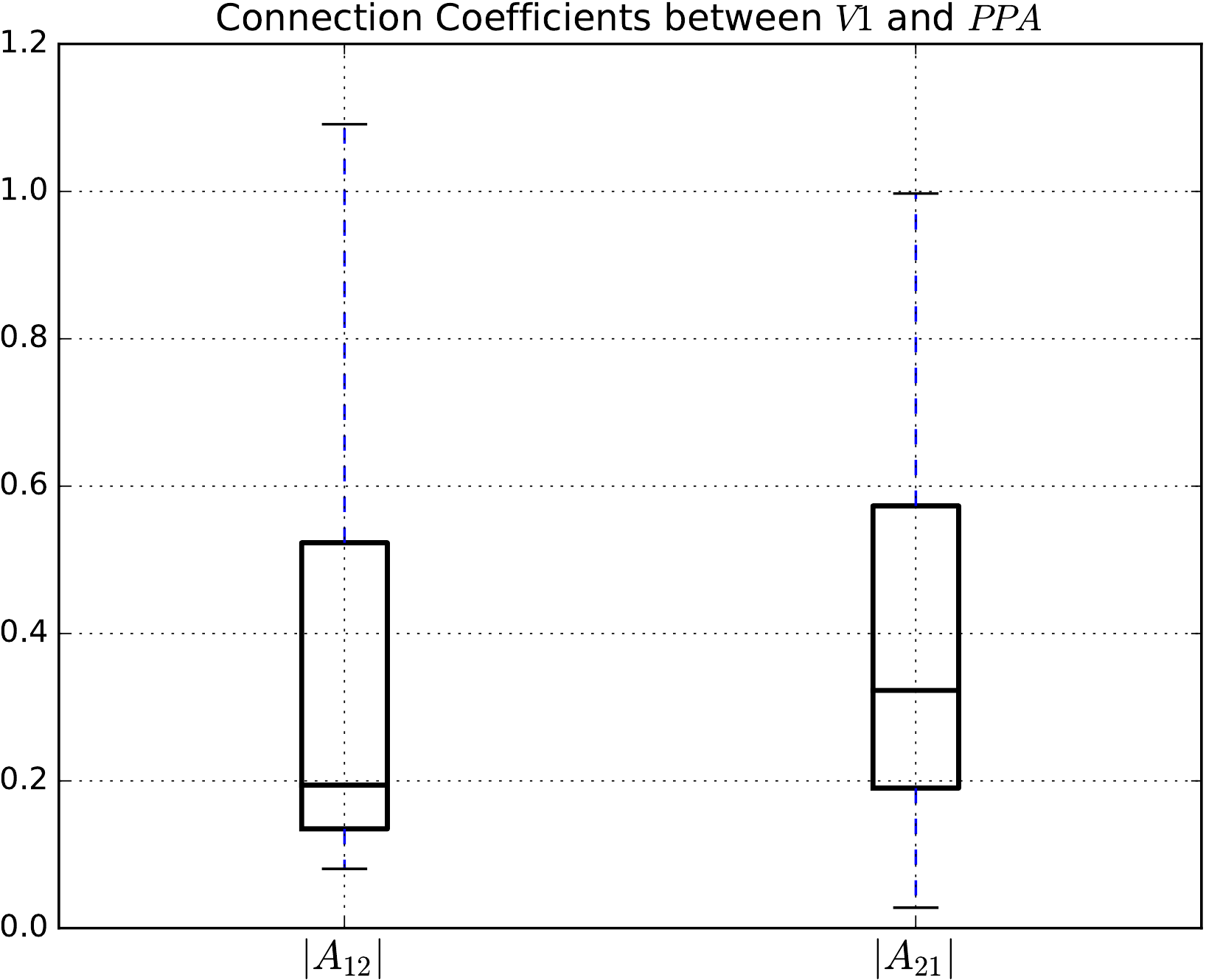}
                \caption{V1 and PPA under resting state.}
                \label{fig:vp_s}
        \end{subfigure}
        ~ %add desired spacing between images, e. g. ~, \quad, \qquad, \hfill etc.

        \begin{subfigure}[b]{0.5\textwidth}
        		\centering
        		\includegraphics*[width=0.75\textwidth]{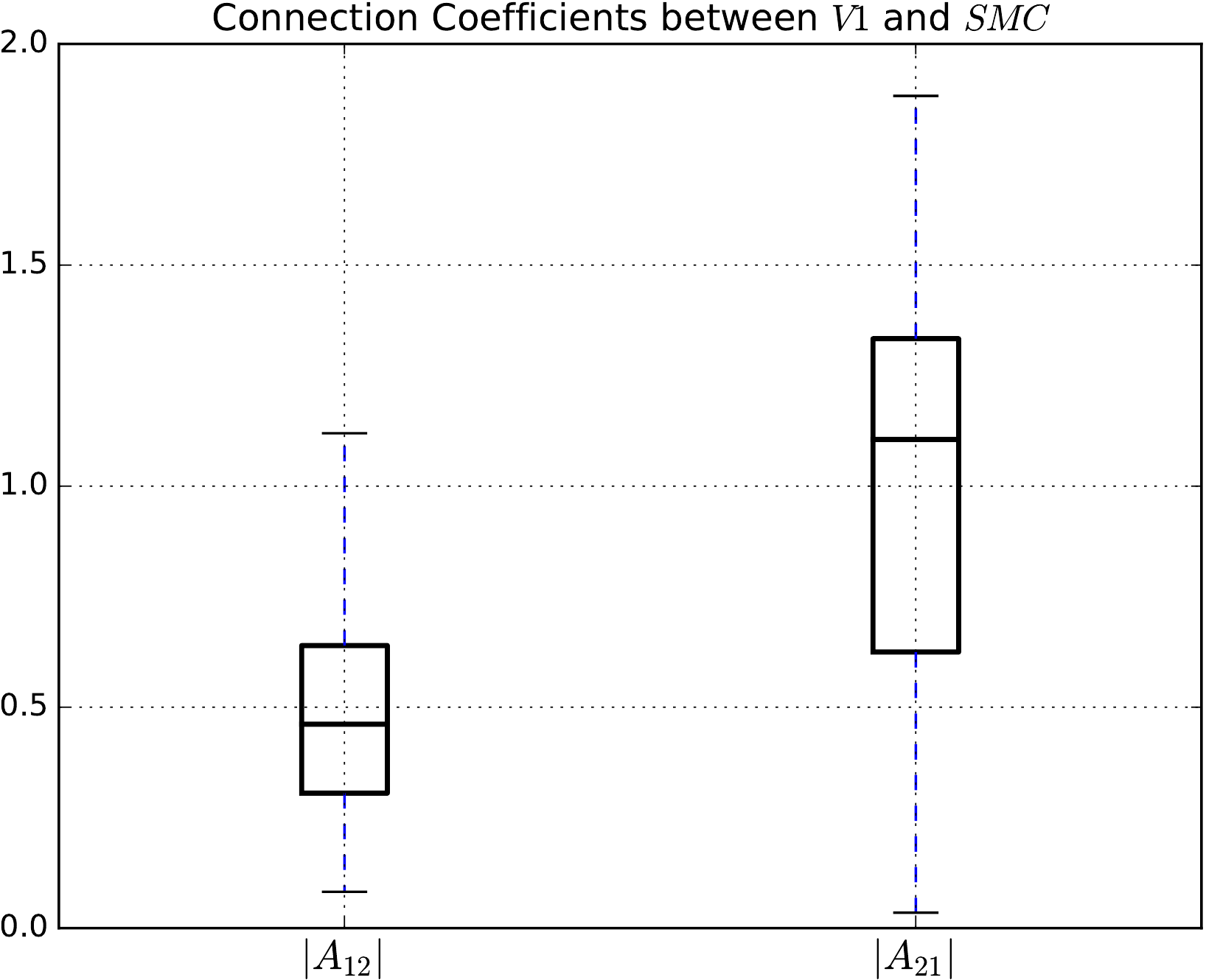}
        		\caption{V1 and SMC under stimulus based state.}
        		\label{fig:vs}
        \end{subfigure}
        
        \begin{subfigure}[b]{0.5\textwidth}
        		\centering
        		\includegraphics*[width=0.75\textwidth]{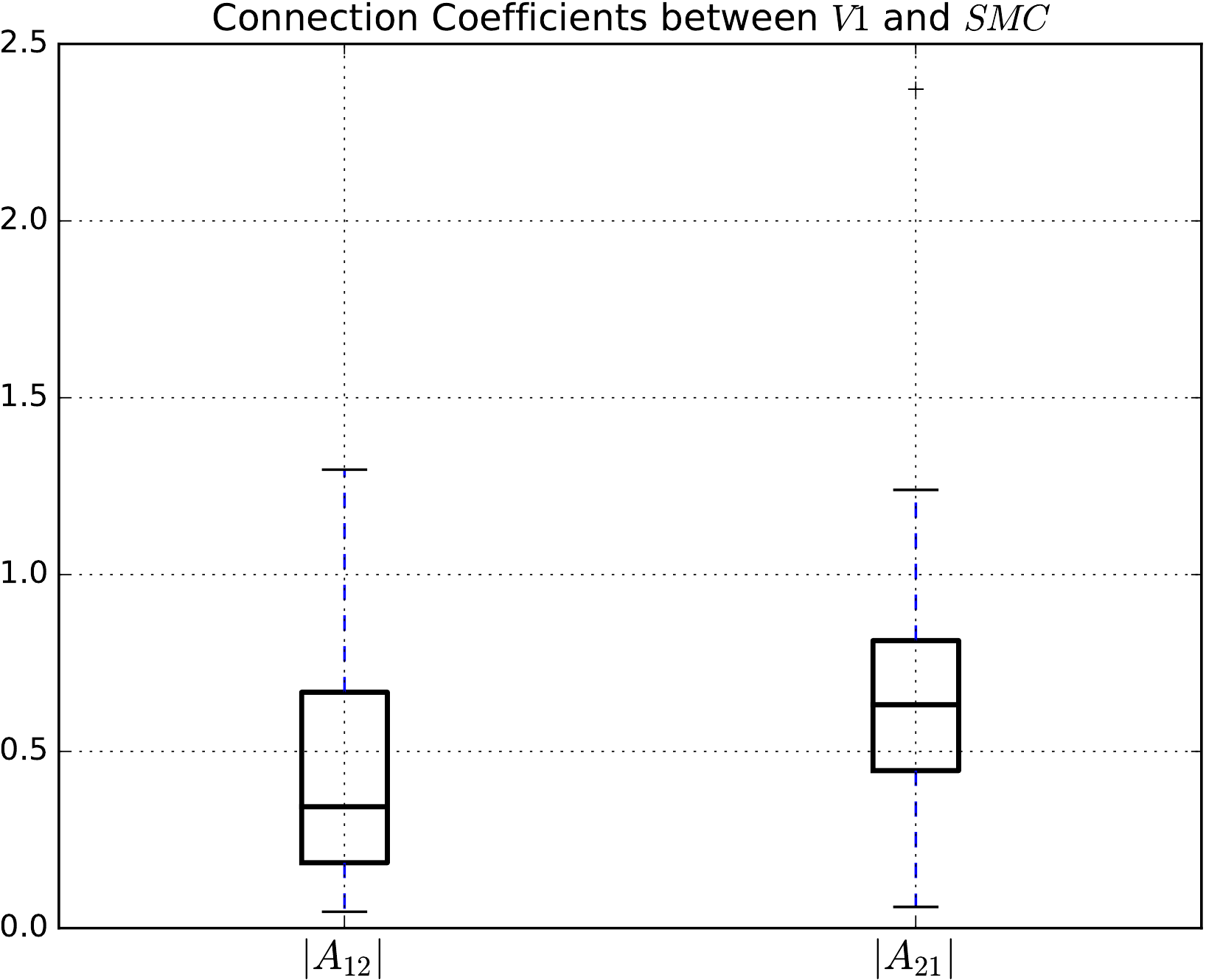}
        		\caption{V1 and SMC under resting state.}
        		\label{fig:vs_s}
        \end{subfigure}
        \caption{Estimations result of DDCM with the experimental fMRI data.}
        \label{fig:fmri}
\end{figure}

The simulation result of V1 and PPA under both resting and stimulus based states are shown in Figure (\ref{fig:vp}) and (\ref{fig:vp_s}). It can be seen that under the resting state, V1 does not exhibit a dominating influence over PPA. However, under the stimulus based state, $|A_{21}|$ is increased considerably compared to $|A_{12}|$. In other words, V1 shows stronger influences over PPA as expected. Figure (\ref{fig:vs}) and (\ref{fig:vs_s}) have shown a similar pattern for the regions V1 and SMC. The result is consistent with the expectations and our previous result using DI~\cite{di}.

\section{Conclusions}

This paper investigated the discrete time DCM (DDCM) and its relationship with Directed Information (DI) and Granger Causality (GC). Based on information theory, we revealed the conditional equivalence between DDCM and DI in characterizing the causal relationship between two brain regions. The theoretical techniques were demonstrated using fMRI data obtained under both resting state and stimulus based state. Our numerical analysis was consistent with that reported in previous study.

\bibliographystyle{IEEEtran}
\bibliography{reference}
%\nocite{*}

\end{document}